\documentclass[twoside,reqno]{HERON}
\usepackage{epsfig,cite,colordvi}
\usepackage{graphicx}
\usepackage{url}
\usepackage{amssymb,amsmath,amscd,epsf}
\usepackage{times}
\usepackage{makeidx}
\usepackage{braket}
\usepackage{slashed}
\usepackage{tikz-feynman,contour}
\usepackage{tikz-feynhand}
\usepackage{physics}
\usepackage{mathdots}
\usepackage[normalem]{ulem}

\newcommand{\ud}{{\rm{d}}}
\def\*#1{\mathbf{#1}}

\begin{document}

\title{Nuclear models for inclusive lepton-nucleus scattering in the quasi-elastic region and beyond}

\runningheads{Nuclear models for inclusive lepton-nucleus scattering in the quasi-elastic region and beyond}{V. Belocchi, M.B. Barbaro, A. De Pace, M. Martini}

\begin{start}

\author{V. Belocchi}{1,2}, \coauthor{M.B. Barbaro }{1,2}, \coauthor{A. De Pace}{2}, \coauthor{M. Martini}{3,4}

\index{Belocchi, V.}
\index{Barbaro, M.B. }
\index{De Pace, A.}
\index{Martini, M.}

\address{Dipartimento di Fisica, Università di Torino, 
Via P. Giuria 1, 10125 Torino, Italy}{1}

\address{Istituto Nazionale di Fisica Nucleare,
Sezione di Torino, Via P. Giuria 1, 10125 Torino, Italy}{2}

\address{IPSA-DRII, 63 boulevard de Brandebourg, 94200 Ivry-sur-Seine, France}{3}

\address{Sorbonne Université, Université Paris Diderot, CNRS/IN2P3, 
Laboratoire de Physique Nucléaire et de Hautes Energies (LPNHE), Paris, France}{4}


\begin{Abstract}
High-precision measurements in neutrino oscillation experiments require a very accurate description of the lepton-nucleus scattering process. Several cross-section calculations are available, but important discrepancies are still present between different model predictions.
For the quasi-elastic channel, dominated by one particle-one hole excitations, an overview over several nuclear models - specifically Relativistic Fermi Gas, 
SuperScaling Approach, 
Spectral Function, 
Hartree-Fock 
and Random Phase Approximation 
- is presented and compared with data for electron-nucleus scattering, a very important process for testing theoretical models validity, highlighting the specific features of each approach.
Furthermore an ongoing microscopic calculation of the two particle-two hole excitations contribution to the electromagnetic response 
 is presented, and some preliminary results are shown.
\end{Abstract}
\end{start}

\section{Introduction}

Neutrinos are a powerful tool to explore physics beyond the Standard Model and to explain general features like the matter-antimatter asymmetry. However the very weak interaction those particles experience with matter provide a difficult challenge from the experimental point of view. Two enormous projects are going to be up to the task, DUNE~\cite{DUNE:2016hlj} and HK~\cite{Zalipska:2023gkl}, taking advantages from the important results and experience of numerous previous projects (see  \cite{NuSTEC:2017hzk} for a review).
There are several problems one needs to face in order to reach high-precision measurements, like the 
usage of nuclei as targets. The theoretical description of the charged current quasi-elastic process, a key channel for incident neutrino flavour identification, is affected by uncertainties, and it is strongly dependent on the adopted nuclear and interaction model.
Moreover the incomplete knowledge of the incident neutrino energy leads to a complex vertex reconstruction, reducing the measurement accuracy. An effect of this uncertainty is that the integration over the incident neutrino flux makes harder a detailed analysis of the adopted scheme, hiding some effects and models differences.
Electron-nucleus reactions, which are closely related to neutrino-nucleus reactions, have been widely studied in the past both experimentally and theoretically~\cite{Amaro:2019zos,Benhar:2006wy}. Since in this case the beam energy is known with very good accuracy, these reactions represent an ideal benchmark for nuclear calculations. In this work an investigation on inclusive electron-nucleus scattering is presented, focusing on the Quasi Elastic (QE) channel and showing some results for the two particle-two hole (2p2h) process.

The article is organized as follows. In Sect.2 we outline the formalism and in Sect.3 we provide a brief description of the nuclear models used in this work. In Sect.4 we show some results, to compare 
the strength and weakness of these models, 
and compare our predictions with electron-carbon scattering data. In Sect.5 we draw our conclusions and outline future developments.

\section{Formalism}
Lepton-nucleus scattering is a complex and various process, involving several reaction channels, going from the excitation of collective resonances at low energies to deep inelastic scattering at very high energies.
In this work we focus on two of these processes: QE scattering and the excitation of 2p2h states. 

One of the basic ingredients of the inclusive $(l,l')$ cross section, where $l$ is the incident lepton and $l'$ the outgoing one, is the hadronic tensor
\begin{equation}
    W^{\mu \nu}_A=\sum_X \bra{A}J_A^{\mu \dagger}\ket{X}\bra{X}J_A^\nu\ket{A}\delta^4(P_A+q-P_X)\,,
\label{eq:HT}
\end{equation}
where $A$ is the initial nuclear state, $X$ is a possible hadronic final state, $P_A=(M_A,{\bf 0})$ the 
four-momentum of the nucleus $A$ in its rest frame and $M_A$ its mass, $q=(\omega,\mathbf{q})$ the transferred four-momentum  and  $P_X$ the four-momentum of the hadronic final state.

The inclusive cross section is given by the contraction of the hadronic tensor $W^{\mu\nu}_A$ with the leptonic tensor $L_{\mu\nu}$, which carries the dependence on the leptonic variables. It is customary to express this cross section as a linear combination of nuclear response functions, $R_K$, weighted by leptonic factors $V_K$~\cite{Amaro:2002mj}. 
In the case of inclusive electron scattering, $(e,e')$, due to vector current conservation and azimuthal invariance for the nucleus 
the nuclear responses are two, longitudinal and transverse with respect to the momentum transfer $\bf q$, 
and the cross section reads
\begin{equation}
    \ud^2 \sigma/\ud \omega \ud \Omega_{k'} = \sigma_{Mott} (V_LR_L+V_TR_T)
\end{equation}
where
\begin{equation}
    R_L:=W_A^{00} \qquad R_T:=W_A^{1 1}+W_A^{2 2}\,,
\end{equation} 
$\sigma_{Mott}$ is the Mott cross-section, $\Omega_{k'}$ is  outgoing lepton solid angle and $V_L$ and $V_R$ are the leptonic factors 
(see \cite{Amaro:2002mj} for the explicit expressions).

The QE channel is one of the most studied and best understood processes in lepton-nucleus scattering. The description of this interaction mostly relies on the lepton-nucleon scattering, but considering the nucleon as a bound particle which interacts, both in the initial and final state, with the nuclear environment. The microscopic description of this reaction is a complicated many-body problem, which can be treated only approximately.
In the following we briefly review some of the theoretical schemes that can be used to describe the nuclear dynamics in this reaction.

The 2p2h channels is a purely nuclear phenomenon, in which two interacting nucleons of the target system are involved. For this reason the corresponding current in the hadronic tensor is a two-body operator, that describes how the two nucleons respond to an external probe. The nuclear dynamics is more complicated than in the QE case and the evaluation of this contribution requires a heavy computational effort. In our approach the theoretical base is provided by Effective Field Theory, in which mesons are the mediators of the strong force, hence the corresponding currents are Meson Exchange Currents (MEC).

\section{Models}

 In this Section we shortly outline the nuclear models used in this work.

\subsection{Relativistic Fermi Gas}
A simple model is the Relativistic Fermi Gas (RFG)~\cite{Smith:1972xh 
}, in which the nucleus is considered as a system of non-interacting fermions at zero temperature, correlated only by the Pauli principle. 
In spite of its simplicity, the RFG has the merit of being relativistic, as required to nuclear models in the GeV region of interest for many current neutrino experiments.
It is possible and more realistic to include a binding 
energy that keeps the nucleons together. Nucleons are characterized by a momentum $\bf p$ inside the nucleus, with no preferred orientation. The maximum value of $|\bf p|$ is the Fermi momentum $p_F$, a parameter characterizing the specific nucleus. 

\subsection{Spectral Function}
The spectral function  carries information on the nuclear 
ground state. In this approach, for the QE channel, it is possible to apply the Impulse Approximation (IA), that allows for factorization: the lepton-nucleus cross section is an incoherent sum of lepton-nucleon processes, with some specific prescriptions. In particular the spectral function $P(E,\mathbf{p})$ is the probability density function to find, inside the nucleus $A$, a nucleon $N$ with momentum $\mathbf{p}$ and removal energy $E$, the required amount of energy to promote a bound nucleon to a free particle, namely
\begin{equation}
    P(E,\mathbf{p})=\sum_R |\bra{A}\ket{N,\mathbf{p};R,-\mathbf{p}}|^2 \delta\left(E-m_N+M_A-E_R \right) 
\end{equation}%
where $m_N$ is the nucleon mass and the bracket is the amplitude associated to IA, that is the separation between the nucleon involved in the interaction and the spectator residual nucleus $R$.
In this work we adopted the Rome spectral function~\cite{Benhar:1994hw}, obtained starting from a nuclear shell-model, that accounts for the 80\% of the normalization, plus a Nucleon-Nucleon (NN) correlation part. 
The shell-model describes the nucleon dynamics below the Fermi momentum, in a finite nucleus framework. The shells are broadened through Lorentzian distributions. The NN part is obtained using a correlated basis, and provides the high-momentum tail of the nucleon distribution in momentum space. Spectroscopic factors extracted from $(e,e'p)$ experimental data are introduced to account for the partial occupation of the shells.
Another important feature is that the energy absorbed by the nucleus is taken into account defining $\tilde{\omega}$, the effective exchanged energy $\tilde{\omega}:=\omega+m-E-E_{\mathbf p}$ and so $\tilde q =(\tilde \omega, \mathbf{q})$. This quantity is defined in a way that affects not only the final hadronic phase space, but the hadronic vertex too, via the nucleon form factors, which depend on $q^2$.  

Hence the hadronic tensor is
\begin{equation}
    W^{\mu \nu}_A(q)=\frac{1}{(2\pi)^3}\sum_{i=1}^A \int \ud \mathbf{p} \,\ud E P(E,\mathbf{p})\frac{1}{4E_\mathbf{p}E_{\mathbf{p}+\mathbf{q}}}W^{\mu\nu}_i(p,\tilde{q})
\end{equation}
and the cross section 
reads
\begin{equation}
    \frac{\ud^2 \sigma}{\ud \Omega_{k'} \ud E_{k'}}=\int \ud \mathbf{p}\, \ud E P(E,\mathbf{p})\sum_{i=1}^A \frac{\ud^2 \sigma_i}{\ud \Omega_{k'} \ud E_{k'}}(q,\tilde{q})\,\delta(\tilde\omega+E_{\mathbf{p}}-E_{\mathbf{p}+\mathbf{q}}) 
\end{equation}
where $E_{k'}$ and $\Omega_{k'}$ are  the outgoing lepton energy and scattering angle, respectively.

\subsection{SuperScaling Approach}

The SuperScaling Approach (SuSA) is a generalization of the RFG model. In particular in the RFG it is possible to prove that the nuclear responses 
are proportional to a function, called scaling function, that depends on a single variable $\psi$, called scaling variable, rather than on $\omega$ and $\mathbf{q}$ separately~\cite{Alberico:1988bv}. 
An extended analysis of electron scattering data~\cite{Day:1990mf} has shown that such scaling behavior is respected by nature with very good accuracy for energy transfer below the QE peak, namely $\omega < Q^2/2m_N$, and momentum transfer $q > 400$ MeV/c, approximately. At high $\omega$ scaling is violated due to the contribution of non-QE processes, like resonance production and 2p2h excitations, while at low $q$ scaling violations are associated to collective nuclear excitations.
The term "SuperScaling" indicates the independence of the scaling function not only of the kinematics $(q)$ but also of the specific nucleus $(p_F)$~\cite{Donnelly:1999sw}.

The explicit expression for the scaling function is trivial in the RFG: 
\begin{equation}f_{RFG}(\psi)=\frac{3}{4}(1-\psi^2).\end{equation} However this form is very different from the experimental scaling function, due to the nuclear effects (correlations and final state interactions) which distort the parabolic behavior. In the SuSA approach a realistic scaling function is obtained directly from electron-nucleus scattering data. The main goal of this procedure is to include phenomenologically the nuclear dynamics in the scaling function, defined as the reduced cross section 
\begin{equation}
   f_{SuSA}(\psi)\equiv p_F \times \frac{[\ud^2\sigma/\ud \omega \ud \Omega_{k'}]_{(e,e')}^{exp}}{\overline{\sigma}_{eN}} 
\end{equation}
obtained by dividing the data by the single-nucleon cross section,
relying on the factorization ansatz.
Since scaling violations mainly occur in the transverse channel, in SuSA the longitudinal response is used to construct the scaling function. In the original SuSA model the same $f$ is adopted also in the transverse channel, whereas in the more recent SuSAv2 approach~\cite{Gonzalez-Jimenez:2014eqa} two different scaling functions, $f_L$ and $f_T$, are used, based on the predictions of a relativistic mean field calculation.

\subsection{Hartree-Fock}
The Hartree-Fock (HF) nuclear model considers the nucleus as an ensemble of interacting nucleons. The main idea is to consider nucleons as independent particles that move inside the nucleus experiencing a single particle potential, that arises from the average of the interactions with the others nucleons. Starting from on-shell nucleons and adding the HF potential it is possible to define the particle self-energy $\Sigma_{HF}$, represented, at first order, by tadpole (H) and oyster (F) loop diagrams.  
Inserting  $\Sigma_{HF}$ in the Dyson equation for the nucleon 
propagator 
and solving iteratively one gets the dressed propagator $G_{HF}$ \cite{Walecka:1971}. 
The hadronic tensor can be evaluated using the formalism of the polarization propagator, defined as (Dirac indices are omitted)
\begin{equation}
    \Pi(q)=\int \frac{\ud p}{(2\pi)^4}G_{ph}( p, q)
\end{equation} 
where the \emph{particle-hole propagator}
\begin{equation}
    G_{ph}(p,q)=-iG(p+q)G(p)
    \label{Gph}
\end{equation}
has been introduced. 
In the HF approximation the nucleon propagator $G$ inside Eq.~\eqref{Gph} is the HF one. The hadronic tensor is proportional to the imaginary part 
of the polarization propagator~\cite{Serot:1984ey,Amaro:2002mj}.

In this work the interaction, needed to evaluate the proper self-energy, is described using a non relativistic expansion of a meson exchange Bonn potential, that includes several mesons ($\sigma,\,\omega,\,\pi,\,\rho$) \cite{Machleidt:1987hj,Barbaro:1995ez}. Computations are performed in the infinite nuclear matter framework.

\subsection{Random Phase Approximation}
When a probe reaches a nucleus, it can excite a bound nucleon into a free nucleon, creating a so called particle-hole pair. However, if we consider the whole system, the interaction between nucleons must be taken into account, yielding to a correction in the propagation of the probe inside the nucleus. A well-known scheme is the Random Phase Approximation (RPA) prescription.
RPA is conveniently described in the formalism of the polarization propagator, studying how a p-h pair propagates inside the nucleus. 
 Two different typologies of corrections arise: the direct contribution and the exchange contribution. The former is due to the p-h annihilation and subsequent creation, while the latter includes the interaction between the hole and the particle inside the nucleus.

The so called 'ring approximation' includes only the diagram of the first kind: in this way the perturbation series is resummable and formally yields 
\begin{equation}
    \Pi=\Pi_0+\Pi_0V \Pi \quad \Rightarrow \quad \Pi_{ring}=\frac{\Pi_0}{1-\Pi_0 V} = \frac{\Pi_0}{1-\Pi_1^d/\Pi_0}\,
\end{equation} 
where $\Pi_1^d$ is the \emph{direct} first order contribution.
However the 'fully antisymmetric RPA', here called aRPA, includes the exchange terms too, and the perturbation series is not resummable anymore: in this work the continued fraction expansion is adopted to extract an approximate result. The used expansion is kept up to the first order for computational effort reduction, but the second order was computed to check the convergence of the series, with a positive result~\cite{DePace:1998yx}. Furthermore, in this scheme higher order diagrams are included through an effective interaction in the truncated series (see \cite{DePace:1998yx} for details). At the order $n$ one gets
\begin{equation}
    \Pi_n\simeq \Pi_0\Big(\frac{\Pi_1}{\Pi_0} \Big)^n \quad \Rightarrow \quad \Pi^1_{aRPA}=\frac{\Pi_0}{1-\Pi_1/\Pi_0 }
\end{equation}
where $\Pi_1=\Pi_0 V \Pi_0 + \Pi_{1ex}$, sum of \emph{direct} and \emph{exchange} first order terms.
We computed both approaches, ring and aRPA, using the same potential of the HF model. Moreover, the 'free' polarization propagator $\Pi_0$, that describes the p-h propagation without medium corrections, is evaluated using the RFG and the HF models. Self-energies in the HF scheme show a rather cumbersome behaviour in terms of $\omega$ and $\mathbf{q}$. In order to reduce the computational effort a 'biparabolic approximation' is performed: a parabolic fit is extracted for the hole and particle self-energy separately. 
As shown in \cite{Barbaro:1995ez}, this prescription provides a remarkably good approximation of the exact results. In this work this approximation was applied to evaluate the HF  polarization propagator in the perturbation series for both the mentioned RPA schemes.%

\subsection{Meson Exchange Currents}
A single probe can excite more than one nucleon in a single scattering, producing a two particle-two hole state (2p2h). This occurs when the two nucleons belong to a correlated pair  
and the probe interacts with a two-body Meson Exchange Current (MEC). 
The microscopic description of this process is based on 
the non-Linear $\sigma$ model \cite{Hernandez:2008}, including also phenomenological interaction vertices for the $\Delta$ resonance, that provides the most important contributions. 

In this work we assume that the current is carried by pion exchange, neglecting heavier mesons. 
Moreover, we describe the nuclear ground state as a RFG in order
to simplify the still very cumbersome computations. The expressions for the MEC, not reported here for lack of space,  are available in Refs.~\cite{DePace:2003spn,RuizSimo:2016rtu}. Note that this approach is fully relativistic, as required by the 
high transferred energy values and resonances involved.
As a consequence of the anti-symmetry of the two nucleons wave function, we have two different amplitudes, that yield two kinds of contributions in the 2p2h response, the \emph{direct} and the \emph{exchange}, corresponding respectively to the sum of the two squared amplitudes and the interference between the two amplitudes.

\section{Results}
We now report our predictions for the differential cross section of electron-carbon scattering at different kinematics and compare the results with scattering data. Since the complete computation of the 2p2h contribution is still in progress, in comparing with data we focus on the QE contribution only.
\begin{figure}
\centering
\includegraphics[scale=0.25]{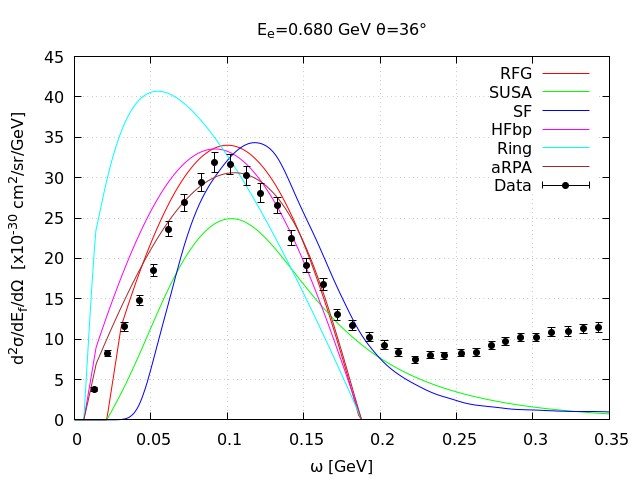}\includegraphics[scale=0.25]{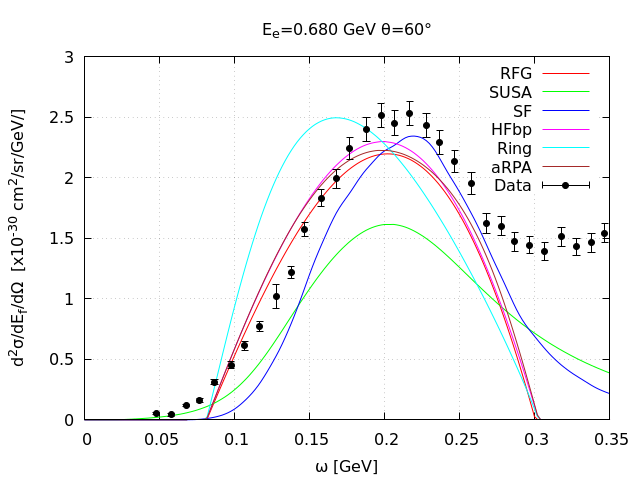}
\includegraphics[scale=0.25]{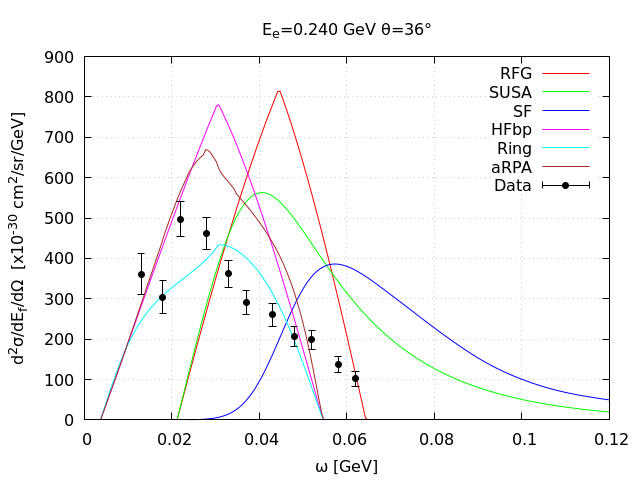}\includegraphics[scale=0.25]{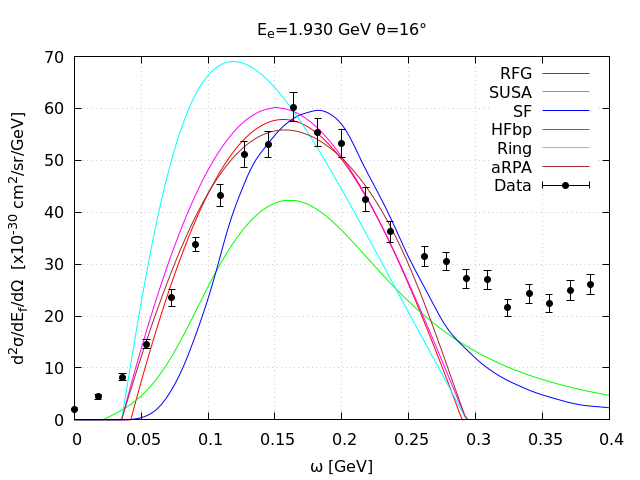}
\caption{Double-differential electron-carbon ($A=12$) cross section plotted versus the energy transfer $\omega$ evaluated with several models (see labels and text) at different electron energies $E_e$ and scattering angles $\theta$ and compared with data~\cite{Benhar:2006er}. The Fermi momentum is $p_F=225$ MeV.}
\label{fig:fig2}
\end{figure}

As it is possible to appreciate in Fig.~\ref{fig:fig2}, the different models behave very differently, and varying also depending on the kinematic region. Before entering into details, some general remarks are in order.
At low energy and momentum transfer Final State Interactions (FSI) and Pauli Blocking (PB) are very important to describe the data in a reliable way. PB is included in every presented model using a step function, except for the SuSA model where Rosenfelder prescription is applied \cite{Rosenfelder:1980nd}. The effect of PB is a reduction of the cross section at very low $\omega$ and $\bf q$ values. FSI are instead crucial to set the peak of the cross section in the right position in $\omega$ space, shifting it towards smaller values. Furthermore, the responses are broadened and display a tail at high energy transfer. The HF and RPA models partially take into account FSI because they include, although approximately, NN interactions in both the initial and final state. SuSA accounts for them since it is based on the extraction of the scaling function from the data, which obviously include all nuclear effects. In the present work the SF does not include FSI, but it is possible to extend the approach by adding a potential to the final outgoing nucleon, using for instance the optical potential theory~\cite{Finelli:2022ckd}.
At high transferred momentum and energy we should expect contributions from non-QE interaction channels, like 2p2h processes and resonances production. So some amount of missing strength is expected when comparing with data.

Due to the ground state description, SuSA and SF show high transferred energy tails in the cross section, partially provided by the NN short range correlations that allow to find off-shell nucleons with momentum higher than $p_F$. For this reason these models have a more realistic shape in this region.
However, both approaches rely on the impulse approximation, valid for $|{\bf q}| > 300 $ MeV, and fail in the low energy and low scattering angles region. 

On the other hand HF, being a mean field model, does not include any NN correlation, while RPA describes long-range correlations. As a consequence, these two models do not produce the high energy tail, but are more efficient in reproducing the low energy region. 

The SuSA approach shows a general lack of strength, due to the fact that the transverse and longitudinal scaling functions are assumed to be equal. As already mentioned, the updated SuSAv2 model, based on a microscopic relativistic mean field calculation, provides a transverse scaling function higher than the longitudinal one~\cite{Gonzalez-Jimenez:2014eqa} which brings the theoretical predictions closer to the data.

The spectral function model provides a realistic description of the nuclear initial state, but the responses suffer severely from the absence of FSI, that affect the peak position and the strength. Better results are achievable adding them \cite{Ankowski:2014yfa}.

HF constitutes a valid model for both the initial and final state of the nucleus, but being based on a mean field does not contain NN correlations.  The RPA scheme, accounting for long-range correlations, produces a redistribution of the strength,  
resulting in a better agreement with data. 
It is worth stressing that the ring approximation and aRPA give very different results at all kinematics, underlining the importance of the exchange contributions in the perturbation series of the polarization tensor. It should be noted that due to the 'biparabolic approximation' (HFbp) in the HF behind the RPA approaches, the cross section range is slightly different from the exact HF. For consistency in Fig.\ref{fig:fig2} the approximated HFbp is shown, so the cross-section energy range is the same.
It's important to mention that the present HF and RPA calculations are performed in nuclear matter and are therefore inadequate to describe the very low energy regime, where finite nucleus effects are important~\cite{Botrugno:2005kn,Pandey:2014tza}.

\begin{figure}
    \centering
    \includegraphics[scale=0.25]{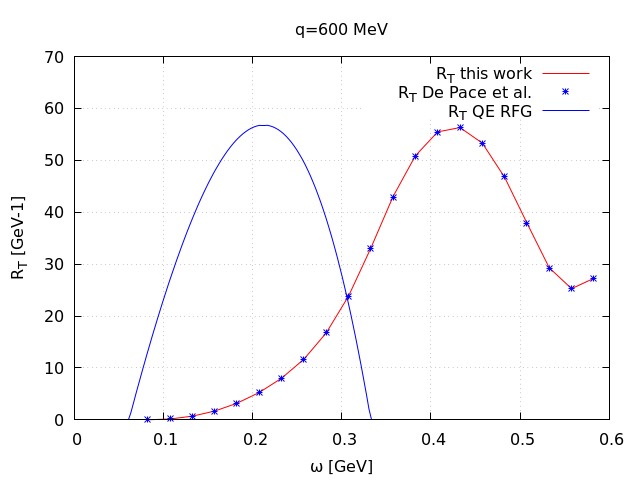}\includegraphics[scale=0.25]{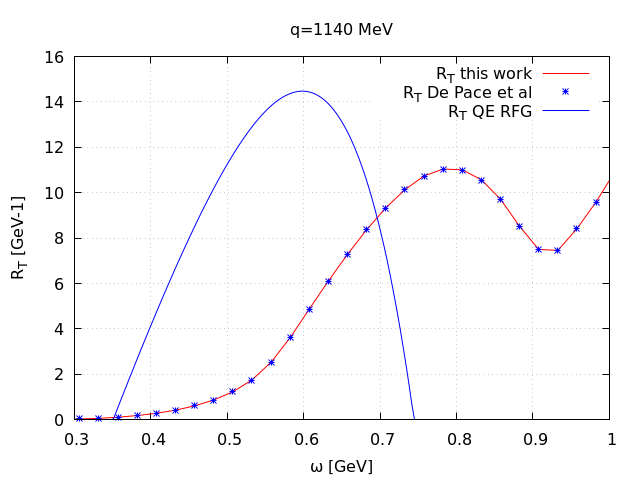}
    \caption{Transverse nuclear response for iron ($A=56$, $p_F$=256 MeV) in the 2p2h channel, with only direct term, shown versus the energy transfer $\omega$ for two different $|\bf q|$ values. The results (solid red) is compared with the one of De Pace {\em et al.}~\cite{DePace:2003spn}. QE transverse response is also represented to see the relative position in energy transfer range with to respect the MEC response. }
    \label{fig:enter-label}
\end{figure}

\begin{figure}
    \centering
    \includegraphics[scale=0.25]{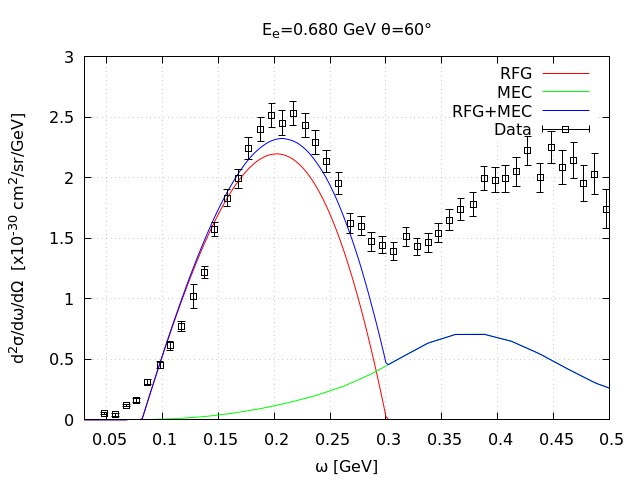}
    \caption{QE and direct-2p2h contributions to electron-carbon cross section shown versus the energy transfer $\omega$ and compared with data~\cite{Benhar:2006er}. The Fermi momentum is $p_F=225$ MeV. }
    \label{fig:MECxsec}
\end{figure}

To conclude, we briefly discuss the results obtained studying the 2p2h process. We proceeded evaluating the nuclear responses in the $\omega$ space, fixing the $\bf{q}$ value.
In Fig.~\ref{fig:enter-label} we report the transverse nuclear responses computed at two given momentum transfer, $|{\bf q}|=600, 1140$ MeV. The calculation is performed for the iron nucleus in order to compare with the results of a previous calculation~\cite{DePace:2003spn}, with which we find perfect agreement. The longitudinal response, not shown here, is negligible.
 It is worth noticing also that the MEC responses as functions of $\omega$ and $|\bf q|$ are truncated when $q$ becomes time-like, and that approaching this limit they start to diverge: this is not a problem because leptonic factors tend to zero at those kinematics, as shown in Fig. \ref{fig:MECxsec}.
It is possible to appreciate the peaked shape of the 2p2h nuclear responses, well defined and separated from the QE peak. Note that the overlap between the two channels increases with increasing $|\bf q|$.

In Fig.~\ref{fig:MECxsec} we present the 2p2h electron-carbon double differential cross section, together with the contribution of QE. For consistency, the QE response is evaluated within the RFG model, also used in the MEC calculation. Note that only the direct contribution to the MEC is included, while the computation for the exchange term is still in progress.
 The MEC contribution is essential to provide strength in the 'dip' region, that is between the QE and $\Delta$-resonance peak. 

\section{Conclusion}

The accurate description of electron scattering data is a necessary validation for nuclear models used in the analyses of neutrino oscillation experiments, which strongly rely on nuclear physics inputs to minimize systematic errors.

With this final goal, we have compared the predictions of different nuclear models - SuperScaling, Spectral Function, Hartree-Fock and Random Phase Approximation - with (e.e') experimental data in the quasi-elastic region, highlighting the success and limits of each model. The Relativistic Fermi Gas results have also be shown as a reference.
A general outcome of the present study is that the HF and RPA approaches are more adequate to describe the low energy transfer region, while the SuSA and SF better describe the high energy tail. 
More specifically, we have analyzed the role of the exchange diagrams in the  RPA approach. It should be noticed that several models based on the ring approximation, largely used for neutrino cross sections~\cite{Martini:2009uj,Nieves:2011pp}, account for antisymmetrization 
adding a phenomenological contact interaction term, in order to reach more reliable results. 
A complete microscopic calculation of the exchange term is very cumbersome and requires an important computational effort. In this work we have accounted for these terms using the continued fraction approximation. The role of antisymmetrization has been shown to be crucial at all the kinematics explored. 

In general, we have found that all approaches provide a reasonably good description of the quasi-elastic peak, except for the very low energy region where finite-nucleus effects, neglected in this work, become dominant.
However, a detailed comparison with experimental cross sections requires the inclusion of non-QE contributions, notably those leading to two particle-two hole excitations.
We have shown preliminary results on the contribution of meson exchange currents to inclusive (e,e') scattering, which have been validated with previous calculations. 

The present work represents a first step towards various developments.
First, the same models can be applied to the study of neutrino-nucleus cross sections and the results compared with the recent data from neutrino experiments (T2K, MicroBooNE, MINERvA). Many studies have been carried out on this topic, but the specific goal of this work is to perform a consistent comparison between existing models, using the same form factors, parameters and code, allowing us to better isolate genuine nuclear model effects. Furthermore, computations for electron-weak processes with aRPA constitutes a new contribution to the field, since in past RPA approaches the exchange terms  was taken into account only with phenomenological prescriptions. Second, the study, up to now limited to the description of inclusive observables, can be extended to the analysis of semi-inclusive scattering, both for the QE and for the 2p2h responses. This kind of data involve not only the leptonic but also the hadronic final state and are more sensitive to nuclear effects, allowing for a better discrimination between different models and approximations. Neutrino physics is moving towards this direction to obtain other ways to extract information on neutrino interactions with nuclei \cite{Franco-Patino:2021yhd}, and a semi-inclusive computation in the 2p2h process is still missing.

\end{document}